\documentclass[a4paper,11pt]{article}
\pdfoutput=1
\usepackage{jheppub} 
\usepackage{orcidlink}
\usepackage{amsmath,amssymb}
\usepackage{graphicx}
\usepackage{float}
\usepackage{soul}
\usepackage[export]{adjustbox}
\usepackage[utf8]{inputenc}
\usepackage{slashed}
\usepackage{bbold}
\usepackage{subfig}
\usepackage{cancel}
\usepackage{subfiles}
\usepackage{subcaption}
\usepackage{amsfonts}
\usepackage{pifont}
\usepackage{xcolor}
\usepackage{physics}
\usepackage{enumitem}
\usepackage{ifthen}
\usepackage{mathtools}
\usepackage{import}
\numberwithin{equation}{section}
\usepackage{standalone}
\usepackage{tikz}
\usepackage{comment}
\usepackage[compat=1.1.0]{tikz-feynman}
\usetikzlibrary{decorations.markings,arrows,arrows.meta,shapes}
\usetikzlibrary{decorations.text}
\usetikzlibrary{bending} 
\DeclareUnicodeCharacter{2212}{-}
\newcommand{\eqs}[2][0.3]{\includegraphics[width=#1\linewidth, valign=c]{#2}}
\def\nn {\nonumber}

\chardef\MyArticleWithColor=\pdfcolorstackinit page direct{0 g}
\graphicspath{{./figures/}}

\title{General finite two-loop amplitude integrand for photoproduction in quark annihilation}

\author[a]{Charalampos Anastasiou}
\author[a]{Julia Karlen\orcidlink{0009-0004-7545-0863}}
\author[a]{Roshni Sahoo\orcidlink{0009-0009-4717-8141}}
\author[b]{George Sterman\orcidlink{0000-0002-7408-8813}}
\author[a]{Matilde Vicini\orcidlink{0009-0003-1914-7328}}
\affiliation[a]{Institute for Theoretical Physics, ETH Zurich,\\8093 Z\"urich, Switzerland}
\affiliation[b]{ C.N.\ Yang Institute for Theoretical Physics and Department of Physics and Astronomy, Stony Brook University,\\Stony Brook NY, 11794-3840 USA}
\emailAdd{babis@phys.ethz.ch}
\emailAdd{karlenj@phys.ethz.ch}
\emailAdd{rsahoo@phys.ethz.ch}
\emailAdd{mvicini@phys.ethz.ch}
\emailAdd{george.sterman@stonybrook.edu}

\abstract{
The construction of integrands free of infrared and ultraviolet singularities may enable the application of numerical methods to evaluate  
loop amplitudes that are inaccessible with analytic techniques. At two loops, finite amplitude integrands have been constructed for the production of off-shell or massive colorless particles via quark annihilation. In this article, we extend this class of processes to include real photons in the final state.

To achieve this, we introduce appropriate momentum flows and counterterms to eliminate singularities that occur because the photons are massless. These singularities arise only locally at the integrand level and do not lead to divergences upon integration. Our treatment also eliminates all power singularities arising from self-energy corrections in the integrand. We extend the analysis to gluon emission from virtual quarks. We believe these new insights will be useful for future extensions of infrared subtraction methods to processes with final-state jets.
}

\begin{document}
\maketitle
\newpage 

\section{Introduction} \label{sec:intro}

Scattering amplitudes for hard-scattering QCD processes factorize into divergent soft and jet functions that are universal, and a process-dependent hard function that is finite \cite{Sen:1982bt, Catani:1998bh, Sterman:2002qn, Dixon:2008gr, Collins:2011zzd, Feige:2014wja, Erdogan:2014gha, Ma:2019hjq}. This factorization offers guidance for developing a subtraction method that renders the process-dependent and finite part of scattering amplitudes in the form of momentum-space integrals over integrands that are locally free of singularities.

Explicit two-loop constructions of finite QCD amplitude integrands have been presented for the production of colorless particles via both quark annihilation~\cite{Anastasiou:2022eym} and gluon fusion~\cite{Anastasiou:2024xvk}. Additionally, Ref.~\cite{Haindl:2025jte} provided key ingredients for a local infrared subtraction method at three loops.
A noteworthy limitation in the quark-annihilation processes studied in Refs~\cite{Anastasiou:2022eym,Haindl:2025jte}, and in analogous QED processes~\cite{Anastasiou:2020sdt}, is the assumption that all final-state particles have a non-vanishing invariant mass. Extending the construction of finite amplitudes to include processes with massless final-state particles is therefore an important step. This publication represents the first advance in that direction.

In this article, we extend the work of Ref.~\cite{Anastasiou:2022eym} to include real photons in the final state. The color structure of processes involving final-state off-shell and on-shell photons is identical. This identity results in the same divergence structure for both types of processes after loop integrations~\cite{Catani:1998bh}. However, at the integrand level, the presence of final-state on-shell photons introduces new types of singularities. We term these singularities {\it transient} because the corresponding integration regions yield finite results without producing $1/\epsilon$ poles (in dimensional regularization).

This effect begins at two loops and can occur whenever a four-, three- or two-point subdiagram carries finite momentum while the remaining loop approaches a soft or collinear region \cite{Sterman:1978bi}.  At fixed values of the subdiagram's loop momentum, $l$, tensor structures 
involving $l$ can form invariants that enhance the singular behavior of the integrand relative to its behavior when $l$ itself approaches a pinched, on-shell region.   

In two-loop amplitudes, transient singularities thus originate from one-loop subgraphs. In the amplitudes we study below, these subgraphs consist of vertex corrections to real photon emission and one-loop self-energy corrections to propagators adjacent to real photons. Self-energy corrections give rise to power-like singularities that cannot be locally tamed at the integrand level by numerator cancellations. 
One-loop vertex corrections contribute terms to the integrand where photons collinear to internal propagators acquire a spurious ``loop polarization"; this phenomenon is analogous to the loop polarizations observed in vertex corrections for collinear virtual gluon emissions, as discussed in Refs.~\cite{Anastasiou:2022eym,Anastasiou:2020sdt}.

In this article, we address these problems by introducing rules to construct two-loop amplitude integrands 
free of transient singularities. Specifically, we eliminate power-like singularities from self-energy corrections through a systematic averaging of the corresponding diagrams over two equivalent loop momentum flows. Furthermore, we eliminate the aforementioned loop polarizations from one-loop vertex corrections by introducing a local vertex counterterm for real photons. This counterterm modifies the integrand but does not alter the amplitude's value, as its integral vanishes when one of the loop momenta is integrated out.

This article is organized as follows. In Section~\ref{sec:Setup}, we describe the amplitude and review the previous construction for off-shell final states from Ref.~\cite{Anastasiou:2022eym}. In Section~\ref{sec:SE_planarQED}, we address the final-state transient singularities arising from self-energy corrections and their removal. Transient singularities 
from loop polarizations associated with triangle corrections to 
on-shell final-state photons are discussed and treated in Section~\ref{sec:triangle_planarQED}. In Section~\ref{sec:amplitude_general}, we construct a general amplitude integrand that is free of infrared and transient singularities for an arbitrary colorless final state. Finally, in Section~\ref{sec:transientgluon}, we discuss how this approach extends to transient singularities in amplitudes with external gluons.

\section{Review and Framework}
\label{sec:Setup}

The results that will be presented in this paper hold for the production of a general colorless final state, which may comprise an arbitrary number of real photons as well as an arbitrary number and type of other massive or off-shell colorless particles. We will show how to construct locally finite integrations for two-loop amplitudes that include the emission of photons. As we shall see, many of our considerations will also apply to the emission of gluons in more general processes.
For illustration, we will focus our discussion to the process of diphoton production at two loops, 
\begin{eqnarray}
    q (p_1) + \bar q (p_2) \to \gamma (q_1) + \gamma (q_2) \, , 
\end{eqnarray}
highlighting the advances that we have made with respect to Ref.~\cite{Anastasiou:2022eym}. As we will not invoke arguments specific to the final-state multiplicity, it will be easy to see that our algorithm  generalizes to any colorless final state. 

The perturbative expansion for the amplitude reads 
\begin{eqnarray}
{\rm M}_{q \bar q \to \gamma \gamma} 
= {\rm M}^{(0)}_{q \bar q \to \gamma \gamma}
+\int \frac{d^Dk}{(2\pi)^D} \, {\cal M}^{(1)}_{q \bar q \to \gamma \gamma}(k) 
+\int \frac{d^Dk}{(2\pi)^D} \,
\frac{d^Dl}{(2\pi)^D} \,{\cal M}^{(2)}_{q \bar q \to \gamma \gamma}(k,l) 
+\ldots 
\end{eqnarray}
As the starting point of our discussion, we will review the one-loop ${\cal M}^{(1)}_{q \bar q \to \gamma \gamma}(k)$ and two-loop ${\cal M}^{(2)}_{q \bar q \to \gamma \gamma}(k,l)$ amplitude integrands as they were constructed for off-shell photons in~\cite{Anastasiou:2022eym}. These integrands were derived with a conventional application of Feynman rules, followed by a coordinated arrangement of loop momentum flows in Feynman diagrams and the addition of counterterms that treated non-local obstructions to the factorization of infrared singularities. We will adopt the notation and conventions of Ref.~\cite{Anastasiou:2022eym} 
as modified slightly in Ref.\ \cite{Anastasiou:2024xvk}.  
The method adopted in Refs.\ \cite{Anastasiou:2022eym} and \cite{Anastasiou:2024xvk} is to shift the standard integrand by a set of integrands that themselves integrate to zero,
\begin{eqnarray}
    &&{\widehat {\cal M}}_{q\bar q \to ew}^{(2),R}(k,l) = {\cal M}_{q\bar q \to ew}^{(2),R}(k,l) - 
    \Delta {\cal M}_{q\bar q \to ew}^{(2),R     }(k,l)\, ,
\label{eq:bfMcalMDeltaM}
\nn\\[4mm]
&& \int d^D k\, \int d^D l \ \Delta {\cal M}_{q\bar q \to ew}^{(2),R}(k,l) = 0\, .
\end{eqnarray}
The superscript $R$ also indicates that these terms are UV regularized by a set of local counterterms.
With these shifts, the integrands become locally factorizable. To be specific, 
\begin{eqnarray}
\label{eq:EWKsubtraction1}
{\cal H}_{q\bar q \to ew}^{(1),R}(k) &=& {\cal M}_{q\bar q \to ew}^{(1),R}(k) - {\cal F}_{q\bar q}^{(1),R}(k) \left[\mathbf{P}_1 \widetilde{\cal M}_{q\bar q \to ew}^{(0)} \mathbf{P}_1 \right] ,\\[2mm]
 \label{eq:EWKsubtraction2}
  {\cal H}_{q\bar q \to ew}^{(2),R}(k,l) &=& {\widehat {\cal M}}_{q\bar q \to ew}^{(2),R}(k,l) - {\cal F}_{q\bar q}^{(2),R}(k,l) \left[\mathbf{P}_1  \widetilde{\cal M}_{q\bar q \to ew}^{(0)} \mathbf{P}_1 \right]
  \nonumber \\
  &\ & \hspace{5mm}
  - {\cal F}_{q\bar q}^{(1),R}(k) \left[ \mathbf{P}_1 \widetilde{\cal H}_{q\bar q \to ew}^{(1),R}(l) \mathbf{P}_1 \right].
\end{eqnarray}
Here, ${\cal H}_{q\bar q \to ew}^{(m)}$ is the $m$-loop hard finite integrand, which includes all process-specific information on the electroweak final state. 
A tilde indicates an integrand without external spinors.  The function ${\cal F}_{q\bar q}^{(m)}$ is the integrand of the $m$-loop form-factor, which absorbs all non-integrable infrared behavior. Both factors include external Dirac spinors. 
The projection matrix factor $\mathbf{P}_1$ is given as, 
\begin{eqnarray}
    \mathbf{P}_1 &\equiv& \frac{\slashed{p}_1\slashed{p}_2}{2\,p_1\cdot p_2}\, .
    \label{eq:p_1-def}
\end{eqnarray}
In this paper, we will construct a set of additional counterterms, which like 
$\Delta {\cal M}_{q\bar q \to ew}^{(2),R}$ in Eq.\ (\ref{eq:bfMcalMDeltaM}) integrate to zero but render the full two-loop integrand locally finite even when on-shell photons are emitted in the final state. The infrared form factors leading to the basic factorization property of Eq.\ (\ref{eq:EWKsubtraction2}) will remain unchanged.

\subsection{Power counting and factorization at one loop}

The leading order amplitude is given by two Feynman diagrams, 
related by Bose symmetry, 
\begin{equation}
{\cal M}^{(0)}_{q \bar q \to \gamma \gamma} 
= \includegraphics[width=0.2\linewidth,valign=c,page=5]{figures/one-loop.pdf}
+\gamma(q_1) \leftrightarrow \gamma(q_2) \, . 
\end{equation}
 The one-loop amplitude consists of the following diagrams, 
\begin{eqnarray}
 \label{eq:M1loop}
{\cal M}^{(1)}_{q \bar q \to \gamma \gamma} 
&=& \includegraphics[width=0.2\linewidth,valign=c,page=4]{figures/one-loop.pdf}
+\includegraphics[width=0.2\linewidth,valign=c,page=3]{figures/one-loop.pdf}
+\includegraphics[width=0.2\linewidth,valign=c,page=6]{figures/one-loop.pdf}
+\includegraphics[width=0.2\linewidth,valign=c,page=2]{figures/one-loop.pdf}
\nonumber \\
&& 
\nonumber \\
&& 
+ \left( \gamma(q_1) \leftrightarrow \gamma(q_2) \right) \, , 
\end{eqnarray}
where the pictured momentum flow and Feynman rules in the Feynman gauge define the integrand. Specifically, the loop momentum of the gluon is always labeled $k$ and is directed from the vertex in the fermion line which is closest to the antiquark ($p_2$) towards the vertex closest to the quark ($p_1$). This momentum flow ensures that, in the collinear limits $k \parallel p_1$ and $k \parallel p_2$, the corresponding singularities factorize locally, at the integrand level. 

To review, the potential for a collinear singularity for the incoming lines can be confirmed by a power counting analysis in the vicinity of either collinear limit.  For example, we study the $k$ integral in the single collinear limit $k\parallel p_1$ in the third and fourth diagrams on the right-hand side of Eq.\ (\ref{eq:M1loop}) by 
expanding $k$ in light cone components adapted to this limit. To do so, we specify a light-like vector $\bar\eta^{(p_1)}$, proportional to $p_1$ and an opposite-moving lightlike vector $\eta^{(p_1)}$ with 
$\eta^{(p_1)}{}^2=\bar\eta^{(p_1)}{}^2=0 $, $\eta^{(p_1)}\cdot \bar\eta^{(p_1)}=1$, with transverse components satisfying $k^{(p_1)}_\perp \cdot \bar\eta_{p_1}= k^{(p_1)}_\perp \cdot \eta^{(p_1)} = 0$. In these terms, the expansion of $k$ is
\begin{eqnarray}
\label{eq:k-p1-expand}
k^\mu = {k \cdot \eta^{(p_1)}  }\, \bar\eta^{(p_1)}{}^\mu +  \, {k \cdot \bar \eta^{(p_1)} }\,  \eta^{(p_1)}{}^\mu +  \, k_\perp^{(p_1)}{}^\mu  \, ,
\end{eqnarray}
In the collinear region $k\parallel p_1$, we scale the components of $k^\mu$ in terms of a variable $\rho$ as
\begin{eqnarray}
\label{eq:k-p1}
    k = \left (k\cdot {\eta}^{(p_1)}\, , k\cdot \bar{\eta}^{(p_1)}\, ,{\bf k}^{(p_1)}_\perp \right ) \sim \left ( \rho^0,\rho^2,\rho \right )\, p_1\cdot \eta^{(p_1)}\, .
\end{eqnarray}
The requirement that the $k$ integration is pinched at the collinear singularity implies that the gluon $k$ flowing up in the diagrams carries a negative fraction of the external energy of $p_1$ \cite{Sterman:1978bj}, with $0\, <\, |k\cdot \eta^{(p_1)}|\,<\, p_1\cdot \eta^{(p_1)}$.
The volume of this collinear $k$ region scales as $\rho^4$.
With this choice of scaling, all terms that appear in denominators $k^2$ and $(k+p_1)^2$ scale as $\rho^2$,
\begin{eqnarray}
\label{eq:co-scaling}
    p_1\cdot k\ 
\sim\ k\cdot \bar{\eta}^{(p_1)}\, k\cdot {\eta}^{(p_1)}\ 
\sim\ k^{(p_1)}_\perp{}^2\ \sim \ \rho^2\, .
\end{eqnarray}
Multiplying denominators of the two collinear lines, $k$ and $k+p_1$ gives $\rho^{-4}$, leading to a logarithmic collinear divergence as $\rho \to 0$. This power counting applies to the third and fourth diagrams on the right of Eq.\ (\ref{eq:M1loop}) in the $k\parallel p_1$ region, with an exactly analogous singular region associated with $p_2$ in the second and fourth. The first self-energy diagram on the right clearly has no infrared singularities. Notice that the two collinear regions are disjoint, except for the soft limit, $k\to 0$.

 The leading numerator factor of the diagrams in Eq.\ (\ref{eq:M1loop}) scales as $\rho^0$ in the $k\parallel p_1$ and $k\parallel p_2$ regions, which does not affect this power counting. However, as $\rho \to 0$, it is easy to verify that in these regions the gluon $k$ carries a longitudinal polarization from the three-point vertex of the initial line, to which it becomes parallel, to the vertex where it is absorbed.
It is precisely the longitudinal polarization of the gluon that leads to factorization via repeated use of the tree-level Ward identity
\begin{eqnarray}
\label{eq:qed-Ward}
    \frac{i}{\slashed{p}}(-i\slashed{k})
    \frac{i}{\slashed{p}+\slashed{k}}
    \ =\ 
    \frac{i}{\slashed{p}}-\frac{i}{\slashed{p}+\slashed{k}}\, ,
\end{eqnarray}
for any momentum $p^\mu$.
The corresponding factorizing singularities appear in the integrated amplitude, and their factorization at the level of the integrand for any electroweak final state is described in Ref.\ \cite{Anastasiou:2022eym}. In that reference, however, final state photons remained massive, eliminating the possibility of similar collinear limits for the outgoing particles.
We now relax that condition and study the possibility of singular behavior when one or more internal lines become collinear to an outgoing on-shell photon.

In the second, third and fourth diagrams on the right-hand side of Eq.~\eqref{eq:M1loop}, one or two real photons are emitted from massless quark propagators in the loop. 
Once we allow the final state photons to go on-shell, the same analysis opens up the possibility of
singularities when $k+p_{1}$ or $k-p_{2}$ becomes collinear with $q_1$ or $q_2$, respectively. For example, in the $k + p_{1} \parallel q_1$ region, we can expand the vector $r^\mu=k^\mu + p_1^\mu$ in the momentum coordinates defined by the photon momentum $q_1$. This expansion is analogous to Eqs.\ (\ref{eq:k-p1-expand}) and (\ref{eq:k-p1}) for the region where $k$ is collinear to $p_1$,
\begin{eqnarray}
\label{eq:r-expand-q1}
    r^\mu = r\cdot \eta^{(q_1)}\, \bar \eta^{(q_1)}{}^\mu\ +\ r \cdot \bar \eta^{(q_1)}\,  \eta^{(q_1)}{}^\mu \ +\ r_\perp^{(q_1)}{}^\mu\ \sim\ 
    \left( \rho^0,\rho^2,\rho \right )\,
    q_1\cdot \eta^{(q_1)}.
\end{eqnarray}
When $q_1^2=0$, this two-to-one transition would give a singularity, by a power counting analysis analogous to Eq.\ (\ref{eq:co-scaling}), if we only consider denominator factors ($\rho^{-4}$) and the volume of integration space ($\rho^{4}$), for $r^\mu$ nearly collinear to $q_1^\mu$. Notice that the region where $k$ is collinear to $p_1$ and the region where $r$ is collinear to $q_1$ are fully disconnected. 

In these one-loop examples, one can easily verify that the numerator vanishes like the scale parameter $\rho$ as we approach the collinear configuration $k+p_1 \parallel q_{1}$.    
The same holds for regions where $k-p_2$ becomes collinear to $q_2$. This suppression is characteristic of the coupling of an on-shell external massless vector to internal massless fermions.

To illustrate the numerator suppression of collinear singularities associated with real photon emissions, we analyze the third diagram as an example. Exhibiting only the relevant fermion propagator and photon-vertex factors of the loop integrand, we have 
\begin{eqnarray}
\includegraphics[width=0.2\linewidth,valign=c,page=1]{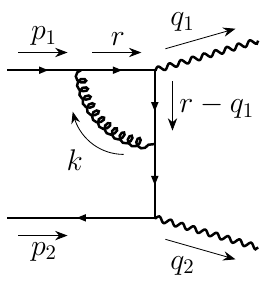} = \ldots  
\frac{(\slashed r - \slashed q_1) \slashed \epsilon_1^* \slashed r }{r^2 (r-q_1)^2} \ldots  
\end{eqnarray}
where, again, $r=k+p_1$. Both of the denominators shown scale as $\rho^2$ in the $r \parallel q_1$ collinear limit. 
However, for a real photon with $q_1^2=0$ and $\epsilon_1^*\cdot q_1=0$, 
the numerator vanishes in this limit due to the transversality of the photon polarization.
We can see this explicitly, 
\begin{eqnarray}
\label{eq:realphotontransversality}
\left. (\slashed r - \slashed q_1) \slashed \epsilon_1^* \slashed r \right|_{r = x q_1} = x \, (x-1) \, \slashed{q_1} \slashed \epsilon_1^* \slashed q_1
= x \, (x-1) \, \left( \slashed{q_1}   2 \epsilon_1^* \cdot q_1  - q_1^2 \slashed{\epsilon}_1^* \right ) =0 \, . 
\end{eqnarray}
Near the singular point, this expression vanishes.
Thus, at one loop, the integrand of Ref.~\cite{Anastasiou:2022eym}, summarized in Eq.~\eqref{eq:M1loop}, can be used for final states with real photons as well.  This suppression is realized locally, and the one-loop amplitude can be integrated numerically~\cite{Kermanschah:2024utt,Capatti:2019edf} through these regions in its standard form.

\subsection{Amplitude integrand at two loops} 

In contrast to the one-loop case, at two loops and beyond we encounter collinear singularities that are power-counting divergent at the integrand level, arising when on-shell photons are emitted in the annihilation of a virtual quark-antiquark pair. We will show that these singularities are transient, meaning that they do not yield divergences after the loop integration is carried out. Nevertheless, such transient singularities do not factorize into universal jet or soft functions. Thus, they prevent naive numerical integration. Our goal in this article will be to remove such non-factoring transient singularities from the two-loop integrand. Before describing how this can be done, we must first review the organization of diagrams presented in Ref.\ \cite{Anastasiou:2022eym}, and the pattern of factorization.

The two-loop amplitude integrand constructed in Ref.~\cite{Anastasiou:2022eym} consists of conventional Feynman diagrams as well as additional diagrammatic terms that ensure IR factorization locally, whenever gluons in loops become collinear to the initial state quark or antiquark. 
Specifically, the loop integrand is a symmetrized function in the loop momenta,
\begin{equation}
\label{eq:M2offshell}
{\cal M}^{(2)}_{q \bar q \to \gamma \gamma}(k,l)     
=\frac{{\cal A}^{(2)}_{q \bar q \to \gamma \gamma}(k,l)+{\cal A}^{(2)}_{q \bar q \to \gamma \gamma}(l,k)}{2}
+ \left( \gamma(q_1) \leftrightarrow \gamma(q_2) \right)\, ,  
\end{equation}
where the diagrams are organized into four categories, 
\begin{eqnarray}
\label{eq:A2decomposition}
{\cal A}^{(2)}_{q \bar q \to \gamma \gamma}(k,l) 
&=& 
{\cal A}^{(2,{\rm fermloop})}_{q \bar q \to \gamma \gamma}(k,l)  + 
{\cal A}^{(2,{\rm islp})}_{q \bar q \to \gamma \gamma}(k,l) +{\cal A}^{(2,{\rm nolp})}_{q \bar q \to \gamma \gamma}(k,l)
\nonumber \\ 
 &&
+\frac{N}{2 C_F} {\cal A}^{(2,{\rm nested})}_{q \bar q \to \gamma \gamma}(k,l-k)
-\frac{1}{2 N C_F} {\cal A}^{(2,{\rm nested})}_{q \bar q \to \gamma \gamma}(k,l)
 \, , 
\end{eqnarray}
The first category, ${\cal A}^{(2,{\rm fermloop})}_{q \bar q \to \gamma \gamma}(k,l)$ contains diagrams with fermion loop subgraphs and all vacuum polarization corrections to a virtual gluon and are treated as in Refs. \cite{Anastasiou:2020sdt, Anastasiou:2022eym}. The construction of the integrand for ${\cal A}^{(2,{\rm fermloop})}_{q \bar q \to \gamma \gamma}(k,l)$ has been discussed in Refs.~\cite{Anastasiou:2020sdt,Anastasiou:2024xvk} and we will not revisit it here as real final-state photons do not induce novel singularities for these diagrams. 
Reference \cite{Kermanschah:2024utt} has presented the numerical evaluation of $N_f$ corrections for generic electroweak production, including real diphoton and triphoton production. Since diagrams in the first class do not generate the transient singularities when outgoing photons are on shell, we will not need to discuss them further. 

In the remaining categories, each diagram has one of three color factors,
\begin{itemize}
    \item $C_FC_A=C_FN$, which we term leading color,
    \item $C_F(C_F-C_A/2)=-C_F/2N$, subleading color, and
    \item $C_F^2$, which, in these terms, is the sum of leading and subleading color.
\end{itemize}
To implement factorization at the local level, it is useful to treat contributions to leading and subleading color separately. In particular, diagrams with the color factor $C_F^2$ are combined with diagrams from the other cases. 

The second category, ${\cal A}^{(2,{\rm islp})}_{q \bar q \to \gamma \gamma}(k,l)$, contains conventional Feynman diagrams that dress the quark-gluon vertices of the initial state with triangle subgraph corrections, or with self-energy corrections to the fermion propagators adjacent to these vertices. 
These graphs give rise to spurious, transient singularities from configurations in which on-shell gluons collinear to $p_1$ and $p_2$ carry a 
polarization proportional to the vertex loop momentum.   Such ``initial-state loop polarizations'' (islp)  can contract with the hard scattering subdiagram. 
Since this polarization is not in general longitudinal, these collinear gluons do not generally factor from the hard scattering. Loop polarizations of this sort thus lead to transient nonfactoring singularities in the sense identified above. 
Reference \cite{Anastasiou:2022eym} combined these diagrams with local counterterms that eliminate the loop polarizations. We have,
\begin{eqnarray}
\label{eq:A2LP}
&& {\cal A}^{(2,{\rm islp})}_{q \bar q \to \gamma \gamma}(k,l) =     
 \eqs[0.13]{./Diags/C1.pdf}
 +\eqs[0.13]{./Diags/C2.pdf}
 +\eqs[0.13]{./Diags/C3.pdf}
 +\eqs[0.13]{./Diags/C4.pdf}
 \nonumber \\ &&
+\eqs[0.13]{./Diags/C5.pdf}
+\eqs[0.13]{./Diags/C6.pdf}
+\eqs[0.13]{./Diags/C7.pdf}
+\eqs[0.13]{./Diags/C8.pdf}
+\eqs[0.13]{./Diags/C9.pdf}
\nonumber \\ &&
+\eqs[0.13]{./Diags/C10.pdf}
+\eqs[0.13]{./Diags/C11.pdf}
+\eqs[0.13]{./Diags/C12.pdf}
+\eqs[0.13]{./Diags/Clp1.pdf}
+\eqs[0.13]{./Diags/Clp2.pdf}
\nonumber \\ &&
+\eqs[0.13]{./Diags/Clp3.pdf}
+\eqs[0.13]{./Diags/Clp4.pdf}\, ,
\end{eqnarray}
where the last four diagrams include the loop polarization counterterms
that eliminate nonfactoring singularities locally for the sum of all preceding diagrams (see Eqs.~(4.38), (4.39), (4.48) and (4.49) of Ref~\cite{Anastasiou:2022eym}).\footnote{For completeness, we note that not all terms in vertex corrections to the incoming quarks generate loop polarizations.}

The third class of diagrams represented in Eq.\ (\ref{eq:A2decomposition}), ${\cal A}^{(2,{\rm nolp})}_{q \bar q \to \gamma \gamma}$, is made up of the following diagrams,
\begin{eqnarray}
\label{eq:A2rest}
&&{\cal A}^{(2, {\rm nolp})}_{q \bar q \to \gamma \gamma}(k,l) =
\eqs[0.13]{./Diags/C21.pdf} 
+\eqs[0.13]{./Diags/C22.pdf}
+\eqs[0.13]{./Diags/C23.pdf}
+\eqs[0.13]{./Diags/C24.pdf}
\nonumber \\ &&
+\eqs[0.13]{./Diags/C27.pdf}
+\eqs[0.13]{./Diags/C28.pdf}
+\eqs[0.13]{./Diags/C29.pdf}
+\eqs[0.13]{./Diags/C30.pdf}
+\eqs[0.13]{./Diags/C33.pdf}
\nonumber \\ &&
+\eqs[0.13]{./Diags/C34.pdf}
+\eqs[0.13]{./Diags/C35.pdf}
+\eqs[0.13]{./Diags/C36.pdf}
+\eqs[0.13]{./Diags/C37.pdf}
+\eqs[0.13]{./Diags/C38.pdf}
\nonumber \\ &&
+\eqs[0.13]{./Diags/C26.pdf}
+\eqs[0.13]{./Diags/C31.pdf}
+\eqs[0.13]{./Diags/C32.pdf}
+\eqs[0.13]{./Diags/C25.pdf}\, .
\end{eqnarray}
This set includes diagrams where the virtual gluons attach to the incoming legs without vertex corrections on the incoming lines. 
These diagrams have only initial state singularities when the outgoing photons go on shell. Moreover, other diagrams in this set, shown in the last line of Eq.~\eqref{eq:A2rest}, where none of the virtual gluons are attached to an initial-state quark or antiquark 
do not develop initial or final singularities.
However, the last diagram of Eq.~\eqref{eq:A2rest}, has spurious singularities associated with the loop momentum of a repeated fermion propagator becoming collinear to arbitrary light-like vectors, generally unrelated to directions of initial or final-state partons. We will discuss this special diagram in Appendix~\ref{sec:nestedSE}. 

The final diagrammatic class, in the second line of Eq.~\eqref{eq:A2decomposition}, consists of planar nested two-loop Feynman diagrams, each with a $C_F^2$ color factor. In this class we only include diagrams with at least one virtual gluon attached to an initial-state quark or antiquark. Explicitly, 
\begin{eqnarray}
\label{eq:A2planQED}
&&{\cal A}^{(2,{\rm nested})}_{q \bar q \to \gamma \gamma}(k,l) =    
\eqs[0.13]{./Diags/C15.pdf}
+\eqs[0.13]{./Diags/C14.pdf}
+\eqs[0.13]{./Diags/C13.pdf}
+\eqs[0.13]{./Diags/C16.pdf}
+\eqs[0.13]{./Diags/C18.pdf}
\nonumber \\ &&
+\eqs[0.13]{./Diags/C17.pdf}
+\eqs[0.13]{./Diags/C19.pdf}
+\eqs[0.13]{./Diags/C20.pdf} \, .
\end{eqnarray}
The diagrams of this class enter the construction of the amplitude integrand in  Eq.~\eqref{eq:A2decomposition} with distinct 
momentum flows for the leading color ($N/2$) and subleading color ($-1/2N$) of the amplitude. This is necessary in order to locally factorize the initial-state collinear singularities, when combined with the second and third class of NNLO diagrams in Eq. (\ref{eq:A2rest}) and Eq. (\ref{eq:A2LP}).

Now we can describe how local factorization occurs for initial-state collinear singularities.
The momentum flows for the subleading color of diagrams of the ``nested" class are shown explicitly in Eq.\ (\ref{eq:A2planQED}). This flow is compatible with 
all crossed, QED-like diagrams in the third class in Eq.\,(\ref{eq:A2rest}). Using only the QED Ward identity in Eq.\ (\ref{eq:qed-Ward}) the sum of all subleading color diagrams, symmetrized in $k\leftrightarrow l$, then factorizes locally, and is rendered finite by the form-factor subtractions of Eq.\ (\ref{eq:EWKsubtraction2}). 

By contrast, the momentum flow for the leading color contribution from the nested diagrams, in which $l$ in (\ref{eq:A2planQED}) is replaced by $l-k$, is compatible with diagrams with a triple-gluon vertex in Eqs. \eqref{eq:A2LP} and (\ref{eq:A2rest}), and their sum factorizes algebraically in both collinear limits after a $k \leftrightarrow l$ symmetrization. Notice that the crossed, QED-like diagrams in Eq.\ (\ref{eq:A2rest}), which are subleading color, play no role in leading color factorization. With the momentum flows shown, local factorization for leading color is purely algebraic and requires both the QED identity of Eq.\ (\ref{eq:qed-Ward}) and the analogous algebraic relation for the triple-gluon vertex,
\begin{eqnarray}
\label{eq:scalar-wi}
    k_\mu \left(2p+k\right)^\mu =\ (p+k)^2 - p^2\, ,
\end{eqnarray}
to which we will return  in our discussion below.
With these identities, the planar ladders factorize algebraically when added to the triple-gluon diagrams. Additional contributions from triple-gluon diagrams factorize among themselves, as shown in Ref.\ \cite{Anastasiou:2022eym}.

In summary, the combination of these four classes of diagrams, as arranged in Eq.~\eqref{eq:A2decomposition}, has the property that the initial state singularities factorize in the integrand. To achieve this property, it is only necessary to add the four loop polarization counterterms of the second class, provide the distinct momentum flows for the leading and subleading contributions of the fourth class. It is not necessary to make subtractions of most diagrams, or to specify finite integrands for 
each diagram individually.
For off-shell photons in the final state, initial state singularities are the only ones that we encounter. 

When we take the photons on-shell, due to the suppression in Eq.~\eqref{eq:realphotontransversality}, there are no final state singularities, associated with virtual quarks becoming collinear to the final-state on-shell photons, in ${\cal A}^{(2, {\rm islp})}_{q \bar q \to \gamma \gamma}(k,l)$ (Eq.~\eqref{eq:A2LP}) and ${\cal A}^{(2, {\rm nolp})}_{q \bar q \to \gamma \gamma}(k,l)$ (Eq.~\eqref{eq:A2rest}). When the photons are on-shell, the nested gluon diagrams ${\cal A}^{(2, {\rm nested})}_{q \bar q \to \gamma \gamma}(k,l)$ of Eq.~\eqref{eq:A2planQED} become singular in final-state collinear regions. This is despite the fact that their integrals remain finite. Our goal is to realize this finiteness at the local level.

These novel singularities have two origins. Nested gluon diagrams with a self-energy correction to the propagator adjacent to an emitted real photon exhibit a local singularity that cannot be suppressed by the vanishing of the numerator, as in Eq.~\eqref{eq:realphotontransversality}, in the collinear limit. 
Nested graphs with a triangle subgraph at the emitted real photon vertex possess a numerator structure different from that of Eq.~\eqref{eq:realphotontransversality}, which does not vanish in the collinear limit if the loop momentum of the subgraph is random. In what follows, we will analyze the form of these final-state singularities for real photon emission, and present methods to eliminate them. 

\section{Self-energy corrections to propagators in nested gluon diagrams} 
 \label{sec:SE_planarQED}

\begin{figure}[H]
    \centering
    \includegraphics[scale=0.7, page=1, valign=c]{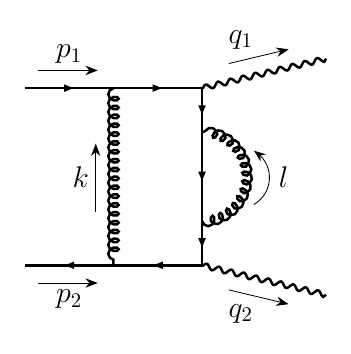}
    \qquad
    \includegraphics[scale=0.7, page=3, valign=c]{figures/self-energy.pdf}
    \qquad
    \includegraphics[scale=0.7, page=6, valign=c]{figures/self-energy.pdf}
    \caption{Diagrams in ${\cal A}^{(2, {\rm nested})}_{q \bar q \to \gamma \gamma}(k,l)$ with self-energy corrections to propagators adjacent to real photons. They have transient final-state singularities. Note that diagrams with self-energy corrections to quark propagators adjacent to the initial states are accounted for separately, in ${\cal A}^{(2, {\rm islp})}_{q \bar q \to \gamma \gamma}(k,l)$.}
    \label{fig:planQEDself-energy}
\end{figure}

In this section, we address the issue of final-state singularities arising from self-energy corrections in the nested gluon diagrams. These diagrams are shown in Fig.~\ref{fig:planQEDself-energy}. 

To illustrate the singularity, let us focus on one of the diagrams, 
\begin{eqnarray}
\label{eq:rRSR-integrand}
\includegraphics[scale=0.7, page=5, valign=c]{figures/self-energy.pdf}
= \ldots  
\frac{i}{\slashed R}  \, 
{\cal S}(l, R) \, 
\frac{i}{\slashed R}  \, 
\slashed \epsilon_1^* \,   
\frac{i}{\slashed r } 
\ldots \,,  \quad r = k+p_1 \,,  R= k+p_1 -q_1 \,, 
\nonumber\\
\end{eqnarray}
where we show only the factors of the diagram adjacent to 
the photon ($q_1$) vertex emerging from the loop,  the fermion propagators in the $k$ loop and the self-energy subgraph, 
\begin{equation}
{\cal S}(l, R)   \equiv  \gamma^\alpha  
\frac 1 {l^2} \frac{1}{\slashed l + \slashed R}
\gamma_\alpha \,,   
\end{equation}
without color factors or other constants.

We study the $r$ integral in the single collinear limit $r\parallel q_1$,  for fixed $l$, expanding $r$ in light cone components analogous to  the one-loop case in Eq.\ (\ref{eq:r-expand-q1}), by introducing light-like vector $\bar\eta^{(q_1)}$, proportional to $q_1$, and an opposite-moving lightlike vector $\eta^{(q_1)}$.
In this collinear region, the denominators of the integrand of Eq.\ (\ref{eq:rRSR-integrand}) each scale as $\rho^2$, and because there are now three of them, combine to give $1/\rho^6$, while the integration volume scales as $\rho^4$. The numerator suppression identified at one-loop, which is of order $\rho$, is not sufficient to eliminate the singularity.

An analogous problem for self-energy corrections in diagrams belonging to ${\cal A}^{(2, {\rm islp})}_{q \bar q \to \gamma \gamma}(k,l)$ was identified in Refs.~\cite{Anastasiou:2020sdt,Anastasiou:2022eym}. In those diagrams, the self-energies in propagators adjacent to the incoming quarks were simplified using a ``tensor reduction'' procedure, which eliminated one of the two repeated quark propagators. Terms were added to the amplitude, which furnished this simplification, and were combined with tensor reduced parts of triangle subgraphs to form the counterterms appearing in the last four diagrams of Eq.~\eqref{eq:A2LP}.

A tensor reduction of self-energy subgraphs will be an important ingredient for treating real photon final-state singularities. It is instructive to view tensor reduction for self-energies 
as a {\it symmetric integration}. That is, as an averaging of self-energy integrands over two equivalent loop momentum flows. Such an averaging does not change the outcome of loop integrations, 
\begin{eqnarray}
\label{eq:SEaveraging}
\int d^Dl \; {\cal S}(l, R)\ =\
\int d^Dl \ \frac{{\cal S}(l, R)\; +\; {\cal S}(-l-R, R)}{2} \, ,
\end{eqnarray}
and we are permitted to substitute the conventional integrand of self-energies with the one appearing in the integral of the right-hand side of Eq. (\ref{eq:SEaveraging}).
The symmetrized integrand is tensor reduced, 
\begin{eqnarray}
\label{eq:SEsymmetrized}
\frac{{\cal S}(l, R)\; + \; {\cal S}(-l-R, R)}{2}\ =\ \left( 1-\frac D 2  \right) 
\frac 1 {l^2 (l+R)^2} \, {\slashed R}   \, .
\end{eqnarray}
Indeed, we can observe that there is no loop momentum $l$ in the numerator. In addition, this expression has a desired inverse fermion-propagator factor $\slashed R$. When we insert the symmetrized self-energy integrand of Eq.~\eqref{eq:SEsymmetrized} in the integrands of the diagrams in Fig.~\ref{fig:planQEDself-energy}, one of the repeated denominators, corresponding to one of the quark propagators adjacent to the self-energy subgraph, is eliminated and the resulting integrand is free of the final-state photon collinear singularity. 

The diagrams of Fig.~\ref{fig:planQEDself-energy}, in addition to the final-state photon collinear singularities which we can treat with tensor reduction as described above, also possess initial state collinear singularities when $k \parallel p_1$ and $k \parallel p_2$. 
At first, it appears that our tensor reduction is incompatible with the factorization of these initial state singularities.
In fact, in order for the local factorization of initial state singularities to take place, the diagrams of Fig.~\ref{fig:planQEDself-energy} need to be combined with several other diagrams from ${\cal A}^{(2, {\rm nolp})}_{q \bar q \to \gamma \gamma}(k,l)$ in Eq.~\eqref{eq:A2rest}. In Eq.~\eqref{eq:A2decomposition}, the integrands of the diagrams are such that the Ward identity cancellations occur algebraically. If we carry out a tensor reduction of the self-energies of the diagrams in Fig.~\ref{fig:planQEDself-energy} alone, these cancellations will be spoiled.      

We will solve this problem with a simple observation. We will use the fact that tensor reduction is an averaging of integrands which correspond to two loop momentum flows. We will then extend the same averaging of momentum flows to the (parts of) other diagrams, which contribute to Ward identity cancellations. As we have remarked, in the second line of Eq.~\eqref{eq:A2decomposition}, we need to consider  distinct momentum routings for the leading and subleading color parts of the diagrams with self-energy subgraphs in Fig.~\ref{fig:planQEDself-energy}. The loop momentum symmetrization will act separately on the two parts of the color decompositions. In what follows, we will describe each color factor separately. 

\subsection{Factorization at subleading color}\label{sec:SE-subleading}

At subleading color, the sum of self-energy and other diagrams which need to be combined together for the factorization of initial state singularities 
consists of 
\begin{align}
\label{eq:SLCgroupSE}
& {\cal A}^{(2), {\rm SLC}}_{q\bar q \to \gamma \gamma}(k,l) 
=  -\frac{1}{2 \, N \,C_F} 
\left( \includegraphics[scale=0.55, page=1, valign=c]{figures/self-energy.pdf}
    + \includegraphics[scale=0.55, page=3, valign=c]{figures/self-energy.pdf}
    + \includegraphics[scale=0.55, page=6, valign=c]{figures/self-energy.pdf}
    \right. 
\nonumber \\
& \left.  
+ \includegraphics[scale=0.55, page=8, valign=c]{figures/self-energy.pdf}
+ \includegraphics[scale=0.55, page=9, valign=c]{figures/self-energy.pdf}
    \right) 
    + \includegraphics[scale=0.55, page=10, valign=c]{figures/self-energy.pdf}
    + \includegraphics[scale=0.55, page=11, valign=c]{figures/self-energy.pdf} \, .
\end{align}
To eliminate the final-state collinear singularity, as discussed before, we would like to consider the three diagrams of the top line in Eq.~\eqref{eq:SLCgroupSE} with tensor reduced self-energy subgraphs. In terms of momentum flows, this tensor reduction amounts to the loop momentum symmetrization $l \leftrightarrow -l-k-p_1+q_1$. To preserve Ward identity cancellations in the limits where $k \parallel p_1$ and $k \parallel p_2$, we will apply this symmetrization to all the diagrams of Eq.~\eqref{eq:SLCgroupSE}, 
\begin{eqnarray}
\label{eq:SLCsymmSErecipe}
 {\cal A}^{(2), {\rm SLC}}_{q\bar q \to \gamma \gamma}(k,l) 
 \to \frac{
 {\cal A}^{(2), {\rm SLC}}_{q\bar q \to \gamma \gamma}(k,l)
 +{\cal A}^{(2), {\rm SLC}}_{q\bar q \to \gamma \gamma}(k,-l-k-p_1+q_1)
 }
 2 \, .
\end{eqnarray}
We can implement this symmetrization by adding a new contribution to the integrand of  Eq.~\eqref{eq:A2decomposition} which reads, 
\begin{eqnarray}
\label{eq:DeltaAlphaSLC}
 \Delta {\cal A}^{(2), {\rm SLC}}_{q\bar q \to \gamma \gamma}(k,l) 
 = \frac{
{\cal A}^{(2), {\rm SLC}}_{q\bar q \to \gamma \gamma}(k,-l-k-p_1+q_1)
-{\cal A}^{(2), {\rm SLC}}_{q\bar q \to \gamma \gamma}(k,l)
 }
 2    \, .
\end{eqnarray}
This counterterm vanishes upon integration over the loop momentum $l$ and therefore does not change the value of the amplitude.
\subsection{Factorization at leading color}

At leading color, the sum of self-energy and other diagrams which need to be combined together for the factorization of initial state singularities 
consists of 
\begin{eqnarray}
&& {\cal A}^{(2), {\rm LC}}_{q\bar q \to \gamma \gamma}(k,l) 
= \frac{N}{2 \,C_F} 
\left(  \includegraphics[scale=0.55, page=2, valign=c]{figures/self-energy.pdf}
    + \includegraphics[scale=0.55, page=4, valign=c]{figures/self-energy.pdf}
    + \includegraphics[scale=0.55, page=7, valign=c]{figures/self-energy.pdf}
    \right. 
\nonumber \\
&& \left.  
+\includegraphics[scale=0.55, page=8, valign=c]{figures/self-energy.pdf}
+\includegraphics[scale=0.55, page=9, valign=c]{figures/self-energy.pdf}
    \right)
    + \includegraphics[scale=0.55, page=15, valign=c]{figures/self-energy.pdf} 
    + \includegraphics[scale=0.55, page=14, valign=c]{figures/self-energy.pdf} \, .
\nonumber \\ 
\label{eq:LCgroupSE}
\end{eqnarray}
In the last two diagrams with triple gluon vertices we have introduced a new vertex with scalar lines. This is based on the ``scalar decomposition" introduced 
in Section 2.4 of Ref.~\cite{Anastasiou:2024xvk}
and Section 5.3 of  Ref.~\cite{Anastasiou:2022eym}, where we split the triple gluon vertex into three distinct ``scalar"-gluon vertices. We decompose the triple-gluon 
vertex as, 
\begin{equation}
\label{eq:Vggg_decomposition}
     \includegraphics[scale=0.55, page=16, valign=c]{figures/self-energy.pdf}
    =  \includegraphics[scale=0.55, page=17, valign=c]{figures/self-energy.pdf} +  \includegraphics[scale=0.55, page=18, valign=c]{figures/self-energy.pdf} +  \includegraphics[scale=0.55, page=19, valign=c]{figures/self-energy.pdf}\, ,
\end{equation}
where 
\begin{eqnarray}
\label{eq:Vssg}
 \includegraphics[scale=0.55, page=17, valign=c]{figures/self-energy.pdf}
&=& -g_s f_{abc} \eta^{\alpha\beta} (k_1-k_2)^{\gamma} \, . 
\end{eqnarray}
This new Feynman rule corresponds to the tree-level rule for the interaction of a color-octet scalar and a gluon times a metric, which is why we refer to the decomposition as “scalar decomposition”. Note, however, that these are not true scalars, they are introduced solely to graphically split the triple-gluon vertex into three terms. Only the specific term for the last two diagrams in Eq. \eqref{eq:LCgroupSE} is involved in cancellations with the other diagrams in ${\cal A}^{(2), {\rm LC}}_{q\bar q \to \gamma \gamma}(k,l)$.

To tensor reduce the non-factorized self-energy subgraphs, we now need to average over two symmetric momentum flows, obtained by $l - k \leftrightarrow q_1-p_1-l$. 
As in the subleading color contributions, in order to preserve Ward identity cancellations in the limits where $k \parallel p_1$  and $k \parallel p_2$, we will apply this symmetrization to all the diagrams of Eq.~\eqref{eq:LCgroupSE}, 
\begin{eqnarray}
\label{eq:LCsymmSErecipe}
 {\cal A}^{(2), {\rm LC}}_{q\bar q \to \gamma \gamma}(k,l) 
 \to \frac{
 {\cal A}^{(2), {\rm LC}}_{q\bar q \to \gamma \gamma}(k,l)
 +{\cal A}^{(2), {\rm LC}}_{q\bar q \to \gamma \gamma}(k,k-l-p_1+q_1)
 }
 2\, .
\end{eqnarray}
As with Eq.~\eqref{eq:DeltaAlphaSLC}, we also implement this symmetrization by adding a new contribution to the integrand of Eq.~\eqref{eq:A2decomposition},
\begin{eqnarray}
\label{eq:DeltaAlphaLC}
 \Delta {\cal A}^{(2), {\rm LC}}_{q\bar q \to \gamma \gamma}(k,l) 
 =
\frac{
{\cal A}^{(2), {\rm LC}}_{q\bar q \to \gamma \gamma}(k,k-l-p_1+q_1)
-{\cal A}^{(2), {\rm LC}}_{q\bar q \to \gamma \gamma}(k,l)
}
 2\, ,
\end{eqnarray}
which also does not change the value of the amplitude upon integration over the loop momentum $l$.

\subsection{Removing initial state singularities}

The counterterms $\Delta {\cal A}^{(2), {\rm SLC}}_{q\bar q \to \gamma \gamma}(k,l)$ and 
$\Delta {\cal A}^{(2), {\rm LC}}_{q\bar q \to \gamma \gamma}(k,l)$  in Eqs.~\eqref{eq:DeltaAlphaSLC} and ~\eqref{eq:DeltaAlphaLC} are designed to remove transient singularities for final-states with real photons, due to self-energy corrections. They are also designed to yield factorized initial state singularities, in the limits $k\parallel p_1$ and $k\parallel p_2$. In both of these collinear limits, the integrands $\Delta {\cal A}^{(2), {\rm SLC}}_{q\bar q \to \gamma \gamma}(k,l)$ and $\Delta {\cal A}^{(2), {\rm LC}}_{q\bar q \to \gamma \gamma}(k,l)$ can be approximated as
\begin{eqnarray}
&&
\Delta {\cal A}^{(2), {\rm SLC}}_{q\bar q \to \gamma \gamma}(k,l) + 
\Delta {\cal A}^{(2), {\rm LC}}_{q\bar q \to \gamma \gamma}(k,l)  
\sim  
\Delta {\cal A}^{(2), {\rm IR}}_{q\bar q \to \gamma \gamma}(k,l)  
\end{eqnarray}
where  
\begin{eqnarray}
\Delta {\cal A}^{(2), {\rm IR}}_{q\bar q \to \gamma \gamma}(k,l) &&= 
\frac{{\cal F}_{q \bar q}^{(1)}(k)}{2} \times \,  
\left\{ 
\frac{-1}{2N C_F}
\left( 
 \includegraphics[scale=0.55, page=21, valign=c]{figures/self-energy.pdf}
-
 \includegraphics[scale=0.55, page=20, valign=c]{figures/self-energy.pdf}
\right)
\right.
\nonumber \\ 
&& 
\left. 
+ \frac{N}{2 C_F}\left(
 \includegraphics[scale=0.55, page=22, valign=c]{figures/self-energy.pdf}
-
 \includegraphics[scale=0.55, page=20, valign=c]{figures/self-energy.pdf}
\right)
\right\}\, .
\end{eqnarray}
In this approximation we 
have
a hard function factor 
that consists of
two tree diagrams with self-energy corrections to a quark propagator 
times
the integrand of a one-loop form factor,
for quark annihilation into a neutral scalar $h$, 
\begin{eqnarray}
\label{eq:Fscalar}
{\cal F}_{q \bar q}^{(1)}(k)   &\equiv& 
\frac{ \includegraphics[scale=0.55, page=23, valign=c]{figures/self-energy.pdf}}{  \includegraphics[scale=0.55, page=24, valign=c]{figures/self-energy.pdf}}
=\frac{ ig_s^2 \, C_F \left(4 p_1\cdot p_2 - 4 k\cdot p_1 + 4 k\cdot p_2- D k^2\right) }{k^2 (k-p_2)^2 (k+p_1)^2} \, . 
\end{eqnarray}

\section{Nested two-loop diagrams with triangle subgraphs}
\label{sec:triangle_planarQED}
\begin{figure}[h]
    \centering
    \includegraphics[width=0.2\linewidth,valign=c, page=4]{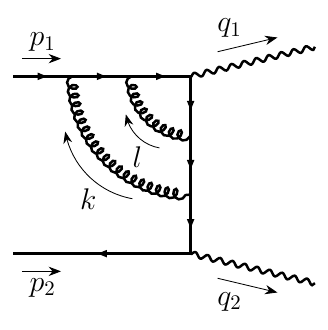}
    \qquad
    \includegraphics[width=0.2\linewidth,valign=c, page=9]{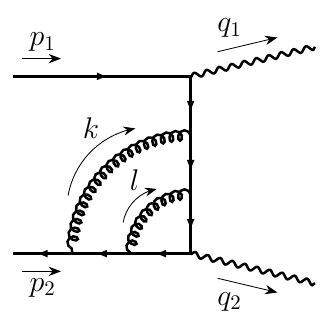}
    \qquad
    \includegraphics[width=0.2\linewidth,valign=c, page=7]{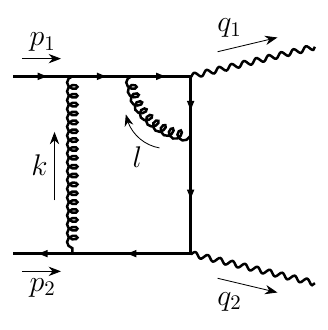}
    \qquad
    \includegraphics[width=0.2\linewidth,valign=c, page=13]{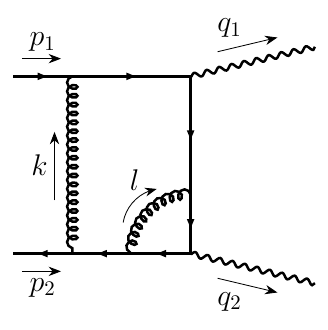}
    \caption{Two-loop diagrams in ${\cal A}^{(2, {\rm nested})}_{q \bar q \to \gamma \gamma}(k,l)$ with one-loop  corrections to real photon vertices. They have a transient final-state singularity.}
    \label{fig:planQEDtriangle}
\end{figure}

We now turn our attention to nested diagrams with triangle subgraphs, which are
corrections to the vertices of real photons. Diagrams in this class are pictured in Fig.~\ref{fig:planQEDtriangle} and have a transient collinear singularity for a final state with a real photon. 

Focusing on the first diagram in Fig.~\ref{fig:planQEDtriangle} for both leading and subleading color contributions, we find the structure
\begin{eqnarray}
\label{eq:C14}
    && -\frac{1}{2NC_{F}}
    \includegraphics[width=0.3\linewidth,valign=c, page=1]{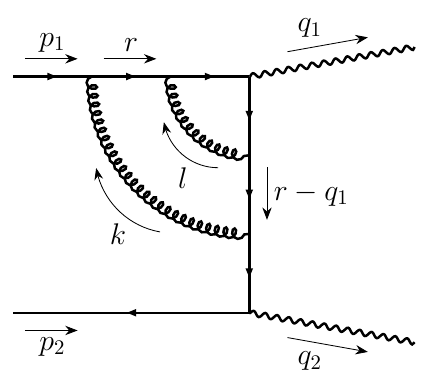} +\frac{N}{2C_{F}}  
    \includegraphics[width=0.3\linewidth,valign=c, page=3]{figures/loop-pol.pdf} \nonumber \\ 
    = 
    && \ldots 
    \frac{i}{\slashed r-\slashed{q_1}}
    \left[ - \frac{1}{2NC_{F}} V_\gamma^{(1)}(r, q_1, l) + \frac{N}{2C_{F}} V_\gamma^{(1)}(r, q_1, l-k) \right] 
    \frac{i}{\slashed r} 
    \ldots \, ,
\quad r = k+p_1 \, ,  
\end{eqnarray}
with
\begin{eqnarray}
\label{eq:Vphoton}
V_\gamma^{(1)}(r, q_1, l) && = -g_s^2 \, e\,  C_F\,  
    \gamma^{\alpha} 
\frac{1}{(\slashed{r} + \slashed{l} - \slashed{q_{1}})} \slashed{\epsilon}_1^{*}(q_{1}) 
\frac{1}{(\slashed{r} + \slashed{l})} 
\gamma_{\alpha} \frac{1}{l^2}\, .
\end{eqnarray}
In the above, we exhibit only the one loop correction $V_\gamma^{(1)}$ 
to the photon vertex and its adjacent propagators, without color or other constant factors. The nested diagrams in Fig.\,\ref{fig:planQEDtriangle} have two color components. According to Eq.~\eqref{eq:A2decomposition}, the leading and subleading color contributions require distinct momentum flows for the loop momentum in the inner subgraph.

Power counting indicates a possible final-state singularity in the limit where both $r=k+p_{1}$ and the loop momentum adjacent to the external photon, $l+r$ or $l-k+r$, become collinear to the photon. 
However, in these double limits, the numerator vanishes similarly as at one loop in Eq.~\eqref{eq:realphotontransversality} and thus these regions of integration are not singular.
 
A second region for which power-counting indicates a potential final-state singularity corresponds to the limit $r=k+p_{1} \parallel q_1$ with the loop momentum $l$ being hard. This turns out to be a true singularity, as in the limit the integrand for the subleading color component becomes, 
\begin{align}
\label{eq:C14limit}
    &\frac{i}{\slashed r-\slashed{q_1}}
    \left( - \frac{1}{2NC_{F}}\right) V_\gamma^{(1)}(r, q_1, l)  
    \frac{i}{\slashed r} \nn\\
    &\xrightarrow{r = x q_1}
    g_s^2 \, e\,\left( - \frac{1}{2N}\right)
    (2-D) \, \slashed q_1 \, 
    \frac{4 x (x-1)}{(r-q_1)^2 \, r^2}
    \frac{l \cdot q_1 \, l \cdot \epsilon_1^*}{l^2 \, (l+r)^2 \, (l+r-q_1)^2}\, .   
\end{align}
The functional form of the $l$ integrand in Eq.~\eqref{eq:C14limit} reveals that the singularity is transient. If one carries out the loop integration, the numerator factors $l \cdot q_1$ and $l \cdot \epsilon_1^*$ can only generate vanishing scalars, $\epsilon^* \cdot q_1 = q_1^2 =0$, in the collinear limit $r \parallel q_1$. Analogously, for the leading color contribution, where the gluon loop carries momentum $l-k$, the integrand in the limit $r\parallel q_{1}$ is the same as Eq. (\ref{eq:C14limit}), after substituting $l$ by $l-k$. We need to remove this transient singularity from the integrand. We will achieve that by adding a counterterm 
that vanishes
upon carrying out the $l$ integration. 

To construct the counterterm, let us first decompose a general momentum $v$ into components parallel to $q_1$, parallel to a light-like vector $\eta_1$ and perpendicular to $q_1 $ and $ \eta_1$, 
\begin{eqnarray}
\label{eq:v-project}
v^\mu &=& \frac{v \cdot \eta_1}{q_1\cdot \eta_1} q_1^\mu
+  \frac{v \cdot q_1}{q_1\cdot \eta_1} \eta_1^\mu + v_{\perp(q_1,\eta_1)}^\mu\, , 
\end{eqnarray}
where the reference vector $\eta_1$ is chosen such that $\epsilon_1^* \cdot \eta_1=0$ and $q_1\cdot\eta_1 \neq 0$. We can then define reflections of the transverse loop momentum as
\begin{eqnarray}
\label{eq:tilde-v-project}
\tilde v^\mu &=& \frac{v \cdot \eta_1}{q_1\cdot \eta_1} q_1^\mu
+  \frac{v \cdot q_1}{q_1\cdot \eta_1} \eta_1^\mu - v_{\perp(q_1,\eta_1)}^\mu\, . 
\end{eqnarray}
Analogous reflections of the loop momenta can be defined with respect to the plane of the second photon momentum $q_{2} $ and a corresponding reference vector $\eta_2$ where $\epsilon_2^*\cdot \eta_2=0$ and $q_2 \cdot \eta_2 \neq 0$. The reflection of the transverse momentum can be defined for the loop momenta $l$ and $k$, as well as for linear combinations. In particular, we require the reflection $\widetilde{l+r}= \widetilde{l}+\widetilde{r}$ of the momenta combination $l+r$ and $\widetilde{l}-\widetilde{k}+\widetilde{r}$, both of which appear as the momentum of a virtual quark in the subleading and leading color contributions, respectively. 

In Ref.~\cite{Anastasiou:2022eym}, loop polarizations were eliminated from triangle subgraph corrections to the incoming states employing a combination of tensor reduction and averaging over loop momentum reflections. We have found that loop momentum reflections can remove loop polarizations which emerge from triangle subgraphs with final-state photons. 
Indeed, we can now define a new vertex $\Delta_{\gamma}^{\mu}(r,q,l,k)$, 
\begin{align}
\label{eq:DeltaTransient}
    & \Delta_\gamma^\mu(r, q, l, k) 
    \equiv 
    \includegraphics[width=0.18\textwidth,page=15,valign=c]{figures/loop-pol.pdf}
     \nonumber \\
     \nonumber \\
    &\equiv g_s^2\, e\, (2-D)\, C_F   \left[ \frac{N}{2C_{F}} \left(  \frac{(l-k+r)^{\mu} (\slashed{l}- \slashed{k}+\slashed{r} )}{(l-k)^{2} (l-k+r)^{2} (l-k+r-q)^{2}}  \right.\right.\nonumber \\
 &\hspace{4.5cm}\left.\left.- \frac{(\widetilde{l-k+r})^{\mu}(\widetilde{\slashed{l}-\slashed{k}+ \slashed{r}})}{(\widetilde{l-k+r} -r )^{2} (\widetilde{l-k+r})^{2} (\widetilde{l-k+r} -q)^{2}} \right)  \right.\nonumber \\ 
 &\hspace{3.5cm}\left. -\frac{1}{2NC_{F}}\, 
     \left(
\frac{(l+r)^\mu \, \left(\slashed{l} + \slashed{r}\right)}{l^2 (l+r)^2 (l+r-q)^2}  
- \frac{\left(\widetilde{l+r}\right)^\mu  \,{\left({\widetilde{\slashed{l}+\slashed{r}}} \right)}}{({\widetilde{l+r} - r})^2 (\widetilde{l+r})^2 (\widetilde{l+r}-q)^2} \right) \right]\, ,\nn\\
\end{align}
where the reflection of the loop momenta is defined with respect to $q$ and a corresponding $\eta$ for which $\epsilon^*(q)\cdot \eta=0.$
This new vertex depends on the incoming quark momentum, the outgoing photon momentum as well as the gluon momenta. The vertex that we introduced here has the property that it vanishes upon integration, 
\begin{eqnarray}
\label{eq:DeltaTransientIntegral}
    \int d^Dl \,  \Delta_\gamma^{\mu}(r, q, l,k) =0 \, , 
\end{eqnarray}
as it is antisymmetric under the change of integration variable $l+r \to \widetilde{l}+\widetilde{r}$ and $l-k+r \to \widetilde{l}-\widetilde{k}+\widetilde{r}$ for the subleading and leading color contribution, respectively. Note that in the limits $l+r\parallel q$ and $l-k+r\parallel q$, the counterterm is finite because the numerators are suppressed by the transversality of the photon polarization.
In addition, when we insert this vertex in between the same propagators that are adjacent to the triangle subgraph of the diagram in Eq.~\eqref{eq:C14} and contract with the photon polarization, we find a collinear limit 
\begin{align}
\label{eq:DeltaCT}
    &\frac{i}{\slashed r-\slashed{q_1}}
    \Delta_\gamma^{\mu}(r, q_1, l,k) \epsilon_{1\mu}^*
    \frac{i}{\slashed r}     \nn\\
    &\xrightarrow[]{r=xq_1}  -g_s^2\, e\, \, (2-D) \, \slashed q_1 \, 
    \frac{4 x (x-1)}{(r-q_1)^2 \, r^2} \nn\\
    & \hspace{1.5cm}  \left(\frac{-1}{2N}\frac{l \cdot q_1 \, l \cdot \epsilon_1^*}{l^2 \, (l+r)^2 \, (l+r-q_1)^2} 
    +\frac{N}{2}\frac{(l-k) \cdot q_1 \, (l-k) \cdot \epsilon_1^*}{(l-k)^2 (l-k+r)^2 (l-k+r-q_{1})^2}  \right)\, .\nn\\
\end{align}
The first term in the above equation is the 
negative
of what we have found in Eq.~\eqref{eq:C14limit} for the triangle subgraph at subleading color in the limit $r \parallel q_{1}$. The property of Eq.~\eqref{eq:DeltaTransientIntegral} constitutes a constructive proof that the singularity of Eq.~\eqref{eq:C14limit} is transient. 

We can now simply use the vertex of Eq.~\eqref{eq:DeltaTransient} as a diagrammatic counterterm, which does not change the value of the integrated amplitude, but has the desired property of removing the transient singularity from the integrand. Indeed, the combination of the following diagrams 
\begin{align}
&\frac{N}{2 C_F} 
\includegraphics[width=0.25\linewidth,valign=c, page=3]{figures/loop-pol.pdf}   
-\frac{1}{2 N C_F} \includegraphics[width=0.25\linewidth,valign=c, page=1]{figures/loop-pol.pdf} + \includegraphics[width=0.25\linewidth,valign=c, page=2]{figures/loop-pol.pdf} \nonumber\\
&\xrightarrow[]{r=xq_1} 0 \, ,
\end{align}
is free of the final-state collinear singularity.

The four diagrams of Fig.~\ref{fig:planQEDtriangle} correspond to the following diagrammatic counterterms, 
\begin{align}
\label{eq:DeltaAlphaNaive}
    &\Delta {\cal A}^{(2),\rm fslp,naive}_{q\bar q \to \gamma \gamma }(k,l)  \nonumber\\
    & = 
    \includegraphics[width=0.2\linewidth,valign=c, page=8]{figures/loop-pol.pdf}
    +\includegraphics[width=0.2\linewidth,valign=c, page=5]{figures/loop-pol.pdf}
    +\includegraphics[width=0.2\linewidth,valign=c, page=14]{figures/loop-pol.pdf}
    +\includegraphics[width=0.2\linewidth,valign=c, page=10]{figures/loop-pol.pdf}
    \,,
\end{align}
where ``fslp" stands for final state loop polarizations. These cancel the final-state collinear singularities. Note that color indices are implicit in the definition of the vertex $\Delta$.

Although final state loop polarization are canceled by the diagrams of Eq.~\eqref{eq:DeltaAlphaNaive}, the same diagrams possess unwanted initial-state singularities locally when loop momentum $k \parallel p_1$ or $k \parallel p_2$. Adding $\Delta {\cal A}^{\rm fslp,naive}_{q\bar q \to \gamma \gamma }$ of Eq.~\eqref{eq:DeltaAlphaNaive} to the amplitude integrand would introduce new singularities that were not originally present (although they integrate to zero).  For true local finiteness, it is necessary to refine the right-hand side of Eq.~\eqref{eq:DeltaAlphaNaive} and render it free of initial state singularities while preserving its final-state singular structure. To satisfy factorization in the $k\parallel p_1$ or $p_2$ limits, Eq.\ (\ref{eq:DeltaAlphaNaive}) would need to be supplemented by diagrams in which gluon $k$ is connected to the internal lines of the composite vertex $\Delta$.  Such diagrams, although singular in initial-state collinear limits, would actually be free of final-state loop polarizations. We can simulate their contributions and restore the factorization of initial-state collinear singularities by a simpler, locally-finite set of counterterms, which integrate to zero.

A revised set of diagrammatic counterterms which achieves these goals is, 
\begin{eqnarray}
\label{eq:DeltaAlpha}
    \Delta {\cal A}^{(2), \rm fslp}_{q\bar q \to \gamma \gamma }(k,l) &=& 
    \includegraphics[width=0.2\linewidth,valign=c, page=8]{figures/loop-pol.pdf}
    +\includegraphics[width=0.2\linewidth,valign=c, page=5]{figures/loop-pol.pdf}
        +\includegraphics[width=0.2\linewidth,valign=c, page=12]{figures/loop-pol.pdf}
    \nonumber \\ 
    &&
    +\includegraphics[width=0.2\linewidth,valign=c, page=14]{figures/loop-pol.pdf}
    +\includegraphics[width=0.2\linewidth,valign=c, page=10]{figures/loop-pol.pdf}
        +\includegraphics[width=0.2\linewidth,valign=c, page=6]{figures/loop-pol.pdf}
    \nonumber \\ 
    &&
    - 
    {\cal F}_{q \bar q}^{(1)}(k) \times \,  
    \left[ 
    \includegraphics[width=0.2\linewidth,valign=c, page=18]{figures/loop-pol.pdf}
    +\includegraphics[width=0.2\linewidth,valign=c, page=17]{figures/loop-pol.pdf}
    \right]
    \, . 
\end{eqnarray}
This refined form differs from the naive counterterm of Eq.~\eqref{eq:DeltaAlphaNaive} by four terms. The third diagram in the first line of Eq.~\eqref{eq:DeltaAlpha} dresses the tree amplitude with two corrections: a loop polarization counterterm $\Delta_\gamma$ on the $q_1$ photon and a factorized one-loop correction on the $q_2$ photon. The $\Delta_\gamma$ on the photon $q_1$ is in between two double lines
that indicate the flow of momentum $k$ in and out of the vertex.
This ensures that the vertex correction in all the diagrams have the same form,
\begin{align}
    \label{eq:Delta-def}
    \includegraphics[scale=0.6,page=16,valign=c]{figures/loop-pol.pdf} = \includegraphics[scale=0.6,page=19,valign=c]{figures/loop-pol.pdf}\, ,
\end{align}
where $r=k+p_1$. This diagram has no final-state collinear singularity, but it is singular in the initial-state collinear limit $k \parallel p_2$.

Analogously, the third diagram in the second line of Eq.~\eqref{eq:DeltaAlpha} dresses the tree amplitude with two corrections: a loop polarization counterterm $\Delta_\gamma$ on the $q_2$ photon with momentum $k$ insertion and a factorized one-loop correction on the $q_1$ photon. This diagram has no final-state collinear singularity, but it is singular in the initial-state collinear limit $k \parallel p_1$. The role of these two diagrams is to add the terms necessary to factorize the initial-state singularities. Unlike what is observed in the sum of the four diagrams of Eq.~\eqref{eq:DeltaAlphaNaive}, a standard analysis of the two initial-state collinear limits reveals that all Ward identity cancellations necessary for factorization occur locally in the sum of the enhanced set of six diagrams in the first and second lines of Eq.~\eqref{eq:DeltaAlpha}.

Finally, the last line of Eq.~\eqref{eq:DeltaAlpha} performs a subtraction of the initial-state singularities which emerged in the sum of diagrams of the first two lines. It is a product of two factors, of which the first is the one-loop form factor for the production of a scalar singlet in quark annihilation of Eq.~\eqref{eq:Fscalar}. The second factor is a contribution to a hard function, dressing the photon vertices of the tree amplitude with a loop-polarization counterterm $\Delta_\gamma^\mu$ with the loop momentum $k$ inserted to match the singular structure of the diagrams in the collinear limits.

\section{Amplitude integrand and hard function remainder for a general colorless final-state}
 \label{sec:amplitude_general}

We can now present a full two-loop amplitude integrand that 
is free of transient final-state collinear singularities and factorizes locally in the infrared limits corresponding to initial state singularities. 
The new integrand generalizes Eq.~\eqref{eq:A2decomposition}, which was 
constructed in Ref.~\cite{Anastasiou:2022eym} for off-shell or massive colorless particles in the final-state. The new, fully locally-finite integrand reads 
\begin{eqnarray}
\label{eq:A2general}
{\cal A}^{(2),{\rm local}}_{q \bar q \to \gamma \gamma}(k,l) 
&=&{\cal A}^{(2)}_{q \bar q \to \gamma \gamma}(k,l)   
+\Delta {\cal A}^{(2),{\rm fslp}}_{q \bar q \to \gamma \gamma}(k,l)
\nonumber \\
&&+\Delta {\cal A}^{(2),{\rm SLC}}_{q \bar q \to \gamma \gamma}(k,l) 
+\Delta {\cal A}^{(2),{\rm LC}}_{q \bar q \to \gamma \gamma}(k,l) 
-\Delta {\cal A}^{(2),{\rm IR}}_{q \bar q \to \gamma \gamma}(k,l)
\end{eqnarray}
It includes the integrand ${\cal A}^{(2)}_{q \bar q \to \gamma \gamma}(k,l)$ constructed with the method of Ref.~\cite{Anastasiou:2022eym} for off-shell and massive colorless final-states. The additional terms have the purpose of removing transient singularities in the spirit of Eq.\ (\ref{eq:bfMcalMDeltaM}), and indeed, by construction, their integrals vanish. 

The $\Delta {\cal A}^{(2),{\rm fslp}}_{q \bar q \to \gamma \gamma}(k,l)$ term,
Eq.\ (\ref{eq:DeltaAlpha}), counters loop polarizations in triangle subgraphs that correct final-state photon vertices. The following terms, $\Delta {\cal A}^{(2),{\rm SLC}}_{q \bar q \to \gamma \gamma}(k,l)$ and $\Delta {\cal A}^{(2),{\rm LC}}_{q \bar q \to \gamma \gamma}(k,l)$ symmetrize self-energies adjacent to photons, as described in Eqs.~\eqref{eq:SLCsymmSErecipe} and \eqref{eq:LCsymmSErecipe}. Their effect is to eliminate doubled propagators, which exacerbate final-state singularities in the integrand. These two terms are constructed in such a way that their initial-state singularities  factorize. The last term, $\Delta {\cal A}^{(2),{\rm IR}}_{q \bar q \to \gamma \gamma}(k,l)$, removes precisely the initial-state singularities of $\Delta {\cal A}^{(2),{\rm SLC}}_{q \bar q \to \gamma \gamma}(k,l)$ and $\Delta {\cal A}^{(2),{\rm LC}}_{q \bar q \to \gamma \gamma}(k,l)$, which were not present in the original integrand.

The full amplitude integrand for the final-state with two real photons is obtained with the symmetrization over the two loop momenta $l,k$, as in Eq.~\eqref{eq:M2offshell},
\begin{equation}
\label{eq:M2onshell}
{\cal M}^{(2),{\rm local}}_{q \bar q \to \gamma \gamma}(k,l)     
=\frac{{\cal A}^{(2),{\rm local}}_{q \bar q \to \gamma \gamma}(k,l)+{\cal A}^{(2),{\rm local}}_{q \bar q \to \gamma \gamma}(l,k)}{2}
+ \left( \gamma(q_1) \leftrightarrow \gamma(q_2) \right)\, .
\end{equation}
In Eq.~\eqref{eq:M2onshell}, we have constructed an integrand that is free of transient singularities and in which all remaining initial state singularities are factorized simultaneously.
These initial state singularities can be removed with the same form-factor subtractions as employed in Ref.~\cite{Anastasiou:2022eym} in Eqs. (7.11)-(7.13). We note that a complete symmetrization as employed above is not completely necessary to achieve local factorization of initial state singularities. Some contributions factorize separately from the 2-loop form-factors (for example the subleading 
color contribution, Eq.\ (\ref{eq:SLCgroupSE})
in Section~\ref{sec:SE_planarQED}). For simplicity and legibility we include the full symmetrization.

So far, we described the construction of the amplitude for the specific process $q \bar q  \to \gamma \gamma $. However, the construction rules can be generalized to any colorless final state, including those containing an arbitrary number of real photons 
in addition to massive electroweak bosons.
 For such a generic state, the integrand is constructed
 by direct generalization
 of Eq.\ (\ref{eq:A2general}), as
\begin{eqnarray}
\label{eq:A2colorless}
{\cal A}^{(2),{\rm local}}_{q \bar q \to {\rm ewk}}(k,l) 
&=&{\cal A}^{(2)}_{q \bar q \to {\rm ewk}}(k,l)   
+\Delta {\cal A}^{(2),{\rm fslp}}_{q \bar q \to {\rm  ewk}}(k,l)
\nonumber \\
&&+\Delta {\cal A}^{(2),{\rm SLC}}_{q \bar q \to {\rm   ewk}}(k,l) 
+\Delta {\cal A}^{(2),{\rm LC}}_{q \bar q \to {\rm     ewk}}(k,l) 
-\Delta {\cal A}^{(2),{\rm IR}}_{q \bar q \to {\rm   ewk}}(k,l),
\end{eqnarray}
where ${\cal A}^{(2)}_{q \bar q \to {\rm ewk}}(k,l)$ is constructed according to the rules presented in Ref.~\cite{Anastasiou:2022eym}. The remaining terms are added to remove transient singularities emerging from real photons in the final state. One-loop corrections to the vertices of outgoing photons acquire a loop polarization in final state collinear limits. These are cured by $\Delta {\cal A}^{(2),{\rm fslp}}_{q \bar q \to {\rm ewk}}(k,l)$, analogous to  $ \Delta {\cal A}^{(2), \rm fslp}_{q\bar q \to \gamma \gamma }(k,l)$ in Eq. (\ref{eq:DeltaAlpha}), where 
\begin{align}
\Delta {\cal A}^{(2),{\rm fslp}}_{q \bar q \to {\rm ewk}}(k,l)
    &= 
    \includegraphics[width=0.25\textwidth,page=1,valign=c]{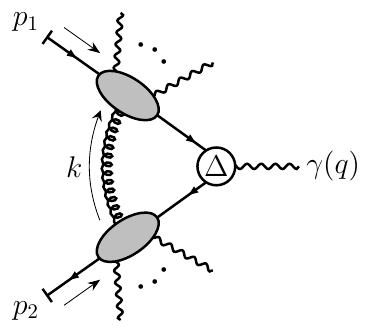}
    +\includegraphics[width=0.25\textwidth,page=2,valign=c]{figures/Julia/tikz_qqani.pdf}
\nonumber \\ 
&\hspace{2cm}
    +\includegraphics[width=0.25\textwidth,page=3,valign=c]{figures/Julia/tikz_qqani.pdf}
-  {\cal F}_{q \bar q}^{(1)}(k) \times 
\includegraphics[width=0.25\textwidth,page=4,valign=c]{figures/Julia/tikz_qqani.pdf}\,.
\label{eq:Dfslp_ewk}
\end{align}
The above expression can be generated from the diagrams of the one-loop amplitude. In every one-loop diagram where a real photon is emitted from the loop, we generate a diagram where the quark-photon vertex is replaced by the vertex $\Delta_\gamma^\mu$, as depicted in the first diagram of Eq.~\eqref{eq:Dfslp_ewk}. To preserve the factorization of initial-state collinear singularities in these graphs, we also introduce diagrams where the loop correction that carries momentum $k$ is graphically disjoint from
$\Delta_\gamma^\mu$,
(second and third diagrams of Eq.~\eqref{eq:Dfslp_ewk}). In the last three graphs, the quark-photon vertex $\Delta_\gamma^\mu(l)$ is positioned outside of the loop that carries momentum $k$, but $k$ is inserted into $\Delta_\gamma^\mu$ via the double lines, as in Eq.\ \eqref{eq:Delta-def}. The last term in Eq.\ \eqref{eq:Dfslp_ewk} is the one-loop form-factor defined in Eq.~\eqref{eq:Fscalar}, multiplied with the tree amplitude, where we replace the vertex of an emitted real photon by $\Delta_\gamma^\mu$ with $k$ momentum insertions.

The transient final state singularities due to self-energies adjacent to photons for subleading color are cured by a direct generalization of Eq.\ (\ref{eq:SLCgroupSE}),
\begin{align}
\label{eq:DSLC_ewk}
    &\Delta {\cal A}^{(2),{\rm SLC}}_{q \bar q \to {\rm   ewk}}(k,l) \nn\\
    &= \frac{1}{2}\left[\frac{-1}{2NC_F}\left(
    \includegraphics[scale=0.5,page=5,valign=c]{figures/Julia/tikz_qqani.pdf}
    -\includegraphics[scale=0.5,page=6,valign=c]{figures/Julia/tikz_qqani.pdf}
    +\includegraphics[scale=0.5,page=7,valign=c]{figures/Julia/tikz_qqani.pdf}
    +\includegraphics[scale=0.5,page=8,valign=c]{figures/Julia/tikz_qqani.pdf}\right)\right. \nn\\
    &\left.
   \hspace{1.5cm}+ 
   \includegraphics[scale=0.5,page=9,valign=c]{figures/Julia/tikz_qqani.pdf}
   +\includegraphics[scale=0.5,page=10,valign=c]{figures/Julia/tikz_qqani.pdf} \right]\nn\\
   &-\frac{1}{2}\left[\frac{-1}{2NC_F}\left(
   \includegraphics[scale=0.5,page=11,valign=c]{figures/Julia/tikz_qqani.pdf}
    -\includegraphics[scale=0.5,page=12,valign=c]{figures/Julia/tikz_qqani.pdf}
    +\includegraphics[scale=0.5,page=13,valign=c]{figures/Julia/tikz_qqani.pdf}
    +\includegraphics[scale=0.5,page=14,valign=c]{figures/Julia/tikz_qqani.pdf}
    \right) \right. \nn\\
    &\left.\hspace{1.5cm}
   + \includegraphics[scale=0.5,page=15,valign=c]{figures/Julia/tikz_qqani.pdf}
   +\includegraphics[scale=0.5,page=16,valign=c]{figures/Julia/tikz_qqani.pdf}    \right]
\end{align}
We do not include diagrams where the self-energy or the triangle vertex correction is next to an incoming leg, since these are already accounted for in the initial-state loop polarization terms ${\cal A}^{(2,{\rm islp})}_{q \bar q \to {\rm ewk}}(k,l)$.
We have not indicated 
exception explicitly in the notation of Eq.~\eqref{eq:DSLC_ewk}. The second diagrams on the first and third lines in Eq.~\eqref{eq:DSLC_ewk} subtract all the diagrams where no photon is adjacent to the self-energy correction. These diagrams have the structure of nested self-energy corrections and are treated separately as described in Appendix \ref{sec:nestedSE}. It is worth noting that the counterterm $\Delta {\cal A}^{(2),{\rm SLC}}_{q \bar q \to {\rm   ewk}}(k,l)$ introduces a symmetrization of the self-energies next to all electroweak final states. Such a symmetrization is not required for self-energies adjacent to massive color-neutral particles, as these do not have transient singularities. Nevertheless, including the counterterm as defined in Eq.~\eqref{eq:DSLC_ewk} is beneficial, since it performs a tensor reduction of the self-energy subgraphs.

Similarly, for the leading-color contribution, generalizing Eq.\ (\ref{eq:LCgroupSE}), we obtain
\begin{align}
\label{eq:DLC_ewk}
    &\Delta {\cal A}^{(2),{\rm LC}}_{q \bar q \to {\rm   ewk}}(k,l) \nn\\
    &=  \frac{1}{2}\left[ \frac{N}{2C_F}\left(
    \includegraphics[scale=0.5,page=17,valign=c]{figures/Julia/tikz_qqani.pdf}
    -\includegraphics[scale=0.5,page=18,valign=c]{figures/Julia/tikz_qqani.pdf}
    +\includegraphics[scale=0.5,page=19,valign=c]{figures/Julia/tikz_qqani.pdf}
    \right.\right.\nn\\
    & \hspace{2cm}\left.\left. +\includegraphics[scale=0.5,page=20,valign=c]{figures/Julia/tikz_qqani.pdf}\right)
   + \includegraphics[scale=0.5,page=21,valign=c]{figures/Julia/tikz_qqani.pdf}
    +\includegraphics[scale=0.5,page=22,valign=c]{figures/Julia/tikz_qqani.pdf}   \right] \nn \\
    &-\frac{1}{2}\left[\frac{N}{2C_F}\left(
    \includegraphics[scale=0.5,page=23,valign=c]{figures/Julia/tikz_qqani.pdf}
    -\includegraphics[scale=0.5,page=24,valign=c]{figures/Julia/tikz_qqani.pdf}
    +\includegraphics[scale=0.5,page=13,valign=c]{figures/Julia/tikz_qqani.pdf}
    +\includegraphics[scale=0.5,page=14,valign=c]{figures/Julia/tikz_qqani.pdf} \right) \right. \nn\\
    & \hspace{1.5cm}\left.
    +\includegraphics[scale=0.5,page=25,valign=c]{figures/Julia/tikz_qqani.pdf}
    +\includegraphics[scale=0.5,page=26,valign=c]{figures/Julia/tikz_qqani.pdf}\right]\, ,
\end{align}
again not including the diagrams accounted for in the initial-state loop polarization amplitude part. The initial-state singularities of Eq.~\eqref{eq:DSLC_ewk}
and Eq.~\eqref{eq:DLC_ewk} are cured by the product of a one-loop form-factor with the one-loop self-energy diagrams:
\begin{align}
\label{eq:DIR_ewk}
    &\Delta {\cal A}^{(2),{\rm IR}}_{q \bar q \to {\rm   ewk}}(k,l) =  {\cal F}_{q \bar q}^{(1)}(k)\times\left[ \frac{-1}{2NC_F}\left(
    +\includegraphics[scale=0.5,page=28,valign=c]{figures/Julia/tikz_qqani.pdf}
    -\includegraphics[scale=0.5,page=27,valign=c]{figures/Julia/tikz_qqani.pdf} 
    \right) \right.\nn\\
    &\left.\hspace{5cm}+  \frac{N}{2C_F}\left( 
    +\includegraphics[scale=0.5,page=29,valign=c]{figures/Julia/tikz_qqani.pdf}
    -\includegraphics[scale=0.5,page=27,valign=c]{figures/Julia/tikz_qqani.pdf} \right) \right]\, .
\end{align}

We should remark that the full integrand of Eq.~\eqref{eq:A2colorless} and the corresponding finite remainder after form-factor subtractions for initial state-singularities, see Eq. \eqref{eq:EWKsubtraction2}, 
does not contain any loop propagators raised to the second power.  As such, the step of integrating out the energy of the loop momenta in formalisms such as Time Ordered Perturbation Theory~\cite{Bodwin:1984hc, Collins:1985ue, Sterman:1993hfp, Sterman:1995fz,Sterman:2023xdj}, Loop-Tree-Duality~\cite{Catani:2008xa, Aguilera-Verdugo:2019kbz, Aguilera-Verdugo:2020set,JesusAguilera-Verdugo:2020fsn,  Runkel:2019yrs, Capatti:2019ypt,Aguilera-Verdugo:2020kzc, Ramirez-Uribe:2020hes, Capatti:2020ytd,Capatti_cLTD_2020,Sborlini:2021owe, TorresBobadilla:2021ivx, Bobadilla:2021pvr, Benincasa:2021qcb, Kromin:2022txz, Ramirez-Uribe:2022sja} and Cross-Free-Family Representation~\cite{Capatti:2022mly, Capatti:2023shz} simplifies.
This step is a prerequisite for recent methods which aim to numerically integrate loop amplitudes in momentum space~\cite{Vicini:2024ecf,Kermanschah:2024utt,Soper:1999xk,Nagy:2006xy,Gong:2008ww,Becker:2010ng,Assadsolimani:2009cz,Becker:2012aqa,Becker:2011vg,Becker:2012bi,Gnendiger:2017pys,Seth:2016hmv,Capatti:2019edf,Capatti:2020ytd,Capatti:2020xjc,TorresBobadilla:2020ekr,Kermanschah:2021wbk,Rios-Sanchez:2024xtv}.

 The integrand of Eq.~\eqref{eq:A2general} still contains ultraviolet singularities. We have implemented a set of ultraviolet subtractions which were proposed in Ref.~\cite{Anastasiou:2022eym}. They were constructed to preserve Ward identity cancellations in the limit where one loop momentum is in the ultraviolet while a second is in an infrared region. After ultraviolet subtractions, we have verified that the integrand of Eq.~\eqref{eq:A2general} is left with the same initial state singularities as the integrand of Eq~\eqref{eq:A2decomposition} for off-shell photons in the final state.

\section{Transient singularities in gluons emitted from quark lines}
\label{sec:transientgluon}

The focus of this work has been to extend the subtraction method for two-loop amplitudes from off-shell or massive electroweak production via quark annihilation to massless photon production. This work already addresses issues central to a subtraction method for massless QCD processes. For example, amplitude integrands for gluon production via quark annihilation are expected to exhibit transient singularities analogous to those for photons. Therefore, the solutions developed in the previous sections should also form part of a future subtraction method for gluon production.

As a concrete example, let us consider one-loop vertex corrections to gluon emissions 
from quark lines,    
\begin{eqnarray}
    \label{eq:corr_extgluon}
    \frac{i}{\slashed{r} - \slashed{q}} V_{\overline{q}gq}^{(1)\text{,c}} \frac{i}{\slashed{r}}
    \equiv 
    \includegraphics[width=0.25\textwidth,page=1,valign=c]{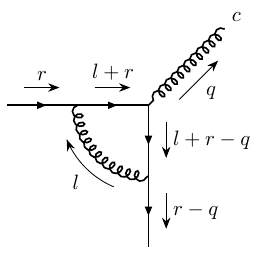}
    +\includegraphics[width=0.25\textwidth,page=2,valign=c]{figures/Julia/tikz_qqani_extgluon.pdf} \, . 
\end{eqnarray}
These one-loop triangles yield transient singularities when they are subgraphs of two-loop diagrams, analogous to the case of photon emission which we treated in Section~\ref{sec:triangle_planarQED}. 
The first contribution above is a QED-like correction, 
\begin{align}
    V_{\overline{q}gq}^{(1)\text{,c,QED}}(r,q,l) &= \frac{1}{2N}\, T^c\, g_s^3 \, \frac{\gamma^\alpha \left(\slashed{r}+\slashed{l}-\slashed{q}\right) \slashed{\epsilon}^* \left(\slashed{r}+\slashed{l}\right) \gamma_\alpha}{l^2(l+r)^2(l+r-q)^2}\, ,
\end{align}
and has exactly the same structure as the subleading color vertex correction in the photon case defined in Eq.~\eqref{eq:Vphoton}. Hence we can use the subleading color part of the vertex $\Delta_\gamma^\mu(r,q,l)$ in Eq.~\eqref{eq:DeltaTransient} to remove the transient singularity. 

For the QCD-like correction to the outgoing gluon in Eq.~\eqref{eq:corr_extgluon}, we have 
\begin{align}
   \includegraphics[width=0.25\textwidth,page=2,valign=c]{figures/Julia/tikz_qqani_extgluon.pdf}  &=  \frac{i}{\slashed{r} - \slashed{q}} V_{\overline{q}gq}^{(1)\text{ c,QCD}} \frac{i}{\slashed{r}}
\end{align}
with 
\begin{align}
    V_{\overline{q}gq}^{(1)\text{ c,QCD}}(r,q,l) &= -\frac{N}{2}\, T^c\, g_s^3 \, \frac{\gamma^\alpha \slashed{l} \gamma^\beta \, \epsilon^*{}^\mu}{l^2(l+r)^2(l+r-q)^2}\nn\\
    &\quad\left( (q-2(l+r))_\mu \, g_{\alpha\beta} + (l+r+q)_\alpha \, g_{\mu\beta} + (-2q+l+r)_\beta \, g_{\alpha\mu}\right) \, .
    \label{eq:VQDC}
\end{align}
Only the first term in the bracket on the second line leads to a transient singularity. Once we multiply the vertex correction by the two fermion propagators $r$ and $r-q$, the second and third term in the bracket in Eq.~\eqref{eq:VQDC} are finite in the limit $r\parallel q$. Including the quark propagators, in the limit we have
\begin{align}
    \frac{i}{\slashed{r} - \slashed{q}} V_{\overline{q}gq}^{(1)\text{ c,QCD}} \frac{i}{\slashed{r}}
    &\xrightarrow[]{r=xq} 
    -\frac{N}{2}\, T^c\, g_s^3 \,(2-D) \frac{2 \epsilon^*\cdot l\, (\slashed{r} - \slashed{q}) \slashed{l} \slashed{r} }{l^2(l+r)^2(l+r-q)^2r^2(r-q)^2} \nn \\
    &= -\frac{N}{2}\, T^c\, g_s^3 \,(2-D) \frac{4(x-1)x }{r^2(r-q)^2}\frac{ l\cdot\epsilon^*\, l\cdot q \,  \slashed{q}}{l^2(l+r)^2(l+r-q)^2}\, ,
\end{align}
which again has the same form as the transient singularity of the subleading-color vertex correction in the photon case. 

We can readily generalize the counterterm vertex of Eq.~\eqref{eq:DeltaTransient} for photons to the non-abelian case of gluon emission. The non-abelian vertex counterterm reads  
\begin{align}
\label{eq:DeltaTransient_gluon}
    & \Delta_{\overline{q}gq}^{\mu, c}(r, q, l)\equiv \includegraphics[width=0.2\textwidth,page=3,valign=c]{figures/Julia/tikz_qqani_extgluon.pdf}
     \nn \\
    &\equiv g_s^3 \left(-\frac{1}{2N}-\frac{N}{2}\right)T^c (2-D) 
     \left( \frac{(l+r)^\mu \, \left(\slashed{l} + \slashed{r}\right)}{l^2 (l+r)^2 (l+r-q)^2}   - \frac{\left(\widetilde{l+r}\right)^\mu  \,{\left({\widetilde{\slashed{l}+\slashed{r}}} \right)}}{({\widetilde{l+r} - r})^2 (\widetilde{l+r})^2 (\widetilde{l+r}-q)^2} \right) \, .\nn\\
\end{align}
In \eqref{eq:DeltaTransient_gluon}, we have attributed the same momentum flow to the two independent components with color factors $\frac{N}{2}$ and $\frac{1}{2N}$. The momentum flows in the corresponding color components of other diagrams, which do not generate transient singularities but participate in the factorization of other infrared singularities, should be attributed accordingly.

\section{Conclusion}

In this paper, we studied the production of general colorless final states in quark annihilation at two-loops. Extending the work of Ref.~\cite{Anastasiou:2022eym} on off-shell or massive final-state particles, we developed a method for eliminating the transient singularities that emerge in the case of real photon production. 
These are local singularities of the integrand from integration regions 
that yield finite contributions, without producing divergences in analytic integration.  

In two-loop photoproduction, a type of transient singularity emerges in diagrams with one-loop quark self-energy corrections that are nested subgraphs within a second loop. For these diagrams, we introduced an averaging over loop momentum flows, which has the effect of a tensor reduction for the self-energy subgraphs. The tensor reduction produces numerator factors in the integrand that locally eliminate the transient singularities.

In order to preserve the Ward identity cancellations at play in the factorization of other, initial-state singularities, the averaging of momentum flows is also applied to diagrams with self-energy corrections that are not nested and do not have transient photon singularities. For these non-nested diagrams, however, the combination of momentum flows is not equivalent to a tensor reduction.

Transient singularities also emerge from nested one-loop vertex corrections to photons. These are a final-state analogue of the loop polarizations from initial-state quarks known from Refs.~\cite{Haindl:2025jte,Anastasiou:2020sdt,Anastasiou:2022eym}. We removed these singularities by introducing a set of counterterm Feynman diagrams that have a new photon-quark-quark vertex. These Feynman diagrams counter the final-state loop polarizations from photons without altering the value of the integrated amplitude, as the new vertex vanishes when one of the loop momenta is integrated. In this article, we also generalize the vertex counterterm to the non-abelian case of a one-loop gluon-quark-quark interaction.

For the subtraction of ultraviolet singularities, we followed the procedure described in Ref.~\cite{Anastasiou:2022eym}. We have explicitly and analytically confirmed that the subtracted two-loop amplitude for diphoton production is finite in all potentially singular infrared and ultraviolet limits.

We look forward to integrating our finite amplitude integrands for generic colorless production processes with numerical methods, such as those developed in Refs.\ 
\cite{Vicini:2024ecf,Kermanschah:2024utt,Soper:1999xk,Nagy:2006xy,Gong:2008ww,Becker:2010ng,Assadsolimani:2009cz,Becker:2012aqa,Becker:2011vg,Becker:2012bi,Gnendiger:2017pys,Seth:2016hmv,Capatti:2019edf,Capatti:2020ytd,Capatti:2020xjc,TorresBobadilla:2020ekr,Kermanschah:2021wbk,Rios-Sanchez:2024xtv}, and to extending our formalism to processes involving jet production.

\section*{Acknowledgments}
This work was supported in part by the National Science Foundation, award PHY-2210533, and the Swiss National Science Foundation, grant 10001706.

\newpage

\appendix
\section{Nested self-energy corrections}
\label{sec:nestedSE}

Diagrams containing two nested self-energy corrections to a massless fermion line, such as the one in Eq.~\eqref{eq:doublebubble}, develop spurious infrared singularities.
\begin{eqnarray}
\label{eq:doublebubble}
\includegraphics[scale=0.7, page=25, valign=c]{figures/self-energy.pdf}
= \ldots \frac{1}{k^2} \gamma^\mu
\frac{i}{\slashed R}  \, 
{\cal S}(l, R) \, 
\frac{i}{\slashed R}  \, \gamma_\mu
\ldots \,,  \quad  R= k+p_1 -q_1 \,.
\end{eqnarray}
The presence of these singularities may seem surprising at first, since the loop subdiagram Eq.\ (\ref{eq:doublebubble}) is a self-energy with a spacelike external momentum, $p_1-q_1$.  The integrations for all such diagrams, however, encounter singularities for real loop momenta, which, however, do not pinch the integrand.
In (\ref{eq:doublebubble}), these singularities arise when the loop momentum $R$ of the repeated fermion propagator becomes collinear to any light-like vector. 
They occur in any two-loop amplitude in massless QCD and are not unique to the processes studied in this article and Refs.~\cite{Anastasiou:2020sdt,Anastasiou:2022eym}. 
The singularities are not pinched and can be avoided with a contour deformation.  However, as these singularities are transient, there is a simpler way to enable numerical integration.

Let us analyze the spurious singularity for the example diagram in Eq.~\eqref{eq:doublebubble}.  
We exhibit only the components of the diagram within the $k$ loop without color factors or other constants and isolate the self-energy subgraph, 
\begin{equation}
{\cal S}(l, R)   \equiv  \gamma^\alpha  
\frac 1 {l^2} \frac{1}{\slashed l + \slashed R}
\gamma_\alpha \, .
\end{equation}
If the momentum $R=k+p_1-q_1$ becomes collinear to an arbitrary light-like vector $\chi$ 
the diagram is singular, which is evident from a power counting analysis. We can scale the components of $R^\mu$ as a function of $\rho$ as 
in previous cases,
\begin{align}
    R^\mu = R\cdot \eta^{(\chi)}\, \bar \eta^{(\chi)}{}^\mu\ +\ R \cdot \bar \eta^{(\chi)}\,  \eta^{(\chi)}{}^\mu \ +\ R_\perp^{(\chi)}{}^\mu\ \sim\ 
    \left( \rho^0,\rho^2,\rho \right )\,
    \chi\cdot \eta^{(\chi)} \,,
\end{align}
where $\bar \eta^{(\chi)}{}^\mu$ is proportional to $\chi$, analogous to the expansion in Eq. \eqref{eq:k-p1-expand}.
In the diagram of Eq. \eqref{eq:doublebubble}, the two quark propagators each contribute a denominator $R^2$, yielding an overall scaling of $\frac{1}{\left(R^2\right)^2} \sim \rho^{-4}$. 
(Contour deformation can be carried out in momentum component $R\cdot\bar\eta^{(\chi)}$.)
The numerator does not yield any suppression.
Because the direction of the vector $\chi$ is arbitrary, the integration volume scales as only $\rho^2$, and the divergence is power-like ($\rho^{-2}$) when $R$ is lightlike.  \footnote{It is worth noting that in the ``soft" limit, where $R\to 0$, the integration volume is again $\rho^4$, and the integral remains power counting finite, because the fermionic propagators behave as $\rho^{-1}$ in this limit.}

Again, the collinear singularity does not correspond to a pinch
and does not yield a divergence upon integration.  The singularity, can be 
softened at the integrand level by performing a symmetric integration or tensor reduction.  
This modifies the self-energy subgraph with gluon loop momentum $l$ as follows,  
\begin{eqnarray}
{\cal S}(l, R) \to \left( 1-\frac D 2  \right) 
\frac 1 {l^2 (l+R)^2} \, {\slashed R}. 
\end{eqnarray}
The matrix product in \ref{eq:doublebubble} then simplifies to 
\begin{align}
\frac{1}{k^2} \gamma^\mu
\frac{i}{\slashed R}  \, 
{\cal S}(l, R) \, 
\frac{i}{\slashed R}  \, \gamma_\mu & \to 
-\left( 1-\frac D 2  \right) 
\frac 1 {l^2 (l+R)^2}   \gamma^\mu\frac{1}{k^2}\frac{1}{\slashed R}\gamma_\mu\,\nn\\
&=
-\left( 1-\frac D 2  \right) 
\frac 1 {l^2 (l+R)^2}  {\cal S}(k, p_1+q_1)\, ,
\end{align}
where in the second step we identify the remaining self-energy correction depending on $k$. Eliminating one of the repeated propagators with $R^{2}$ via the tensor reduction, suppresses the singularity in the collinear limit $R\parallel \chi$. It remains logarithmic, however,  at the local level. It can still be avoided  altogether by a contour deformation.   Alternatively, we can carry out the energy integrals of the diagram, for example in time-ordered or loop-tree formalism.  The resulting three-dimensional integral is guaranteed to be free of singularities, precisely because it describes a fermion self-energy at spacelike external momentum.

Finally, we note that in this article we found reasons for applying tensor-reduction simplification to all one-loop self-energy corrections to massless propagators that appear as nested subgraphs in a second loop.

\bibliographystyle{JHEP}
\bibliography{biblio}
\end{document}